\documentclass[doublecol]{epl2}
\title{Imitating emotions instead of strategies in spatial games elevates social welfare}
\shorttitle{Imitating emotions instead of strategies in spatial games elevates social welfare}

\author{Attila Szolnoki,\inst{1} Neng-Gang Xie,\inst{2} Chao Wang,\inst{2} Matja{\v z} Perc\inst{3}}
\shortauthor{Szolnoki et al.}

\institute{\inst{1}Research Institute for Technical Physics and Materials Science, P.O. Box 49, H-1525 Budapest, Hungary\\
\inst{2}Department of Mechanical Engineering, Anhui University of Technology, Maanshan City 243002, China\\
\inst{3}Faculty of Natural Sciences and Mathematics, University of Maribor, Koro{\v s}ka  cesta 160, SI-2000 Maribor, Slovenia}


\abstract{The success of imitation as an evolutionary driving force in spatial games has often been questioned, especially for social dilemmas such as the snowdrift game, where the most profitable may be the mixed phase sustaining both the cooperative as well as the defective strategy. Here we reexamine this assumption by investigating the evolution of cooperation in spatial social dilemma games, where instead of pure strategies players can adopt emotional profiles of their neighbors. For simplicity, the emotional profile of each player is determined by two pivotal factors only, namely how it behaves towards less and how towards more successful neighbors. We find that imitating emotions such as goodwill and envy instead of pure strategies from the more successful players reestablishes imitation as a tour de force for resolving social dilemmas on structured populations without any additional assumptions or strategic complexity.}

\begin{document}

\maketitle

\section{Introduction}

Societies facing a social dilemma are at risk of failing to uphold wellbeing in their ranks because there exist strong incentives to put success of individuals above that of the society as a whole. It is therefore in the best, although not completely obvious, interest of all if social dilemmas are mitigated or, if at all possible, altogether avoided. Cooperative behavior \cite{axelrod_84} is something of a holly grail when it comes to resolving social dilemmas. To cooperate traditionally means to sacrifice some fraction of personal benefits for elevating social welfare. However, in the face of natural selection, favoring the fittest and the strongest amongst us, the concept quickly becomes misty and the outlook for cooperators to survive murky. Enter evolutionary games \cite{hofbauer_98, nowak_06, sigmund_10}, which are frequently employed to help us reveal and understand the mechanisms and reasons why cooperation nevertheless prevails and is in fact much more common than one could assume. Examples of recent research works aimed towards this direction include \cite{santos_pnas06, hauert_jtb06b, fu_pla07, gomez-gardenes_prl07, rong_pre10, tomassini_ijmpc07, szolnoki_epl09, cremer_njp09, poncela_epl09, dai_ql_njp10, helbing_pre10b, liu_rr_pa10, lee_s_prl11, van-segbroeck_njp11}.

One of the most rewarding observations in recent history related to the resolution of social dilemmas was that spatial reciprocity can maintain cooperative behavior without additional assumptions or mechanism weighing down on defectors \cite{nowak_n92b}. Other well known mechanisms promoting cooperation include kin selection \cite{hamilton_wd_jtb64a}, direct and indirect reciprocity \cite{axelrod_s81}, as well as group \cite{wilson_ds_an77} and multilevel selection \cite{traulsen_pnas06, szolnoki_njp09}. These as well as related mechanism for the promotion of cooperation have been comprehensively reviewed in \cite{nowak_s06}, and there are a number of recent reviews devoted to evolutionary games that capture succinctly recent advances made along this very vibrant avenue of research \cite{szabo_pr07, schuster_jbp08, roca_plr09, perc_bs10}. Focusing on spatial reciprocity, however, one finds that certain social dilemmas are not susceptible to its workings, and that indeed well-mixed conditions may represent a more favorable environment. Hauert and Doebeli \cite{hauert_n04} reported that, especially for the snowdrift game, the promotion of cooperation by means of imitation on structured populations is problematic because the Nash equilibrium is a mixed phase of cooperators and defectors. Consequently, it is advantageous to imitate strategies that are opposite to neighboring strategies, which ultimately leads to a failure of utilizing advantages of spatial reciprocity. Moreover, while some experimental findings question the importance of imitation \cite{grujic_pone10}, others find that humans may imitate even in situations that may decrease their chance of further success \cite{cook_prsb11}, suggesting that such seemingly maladaptive behavior may be due to the inherent evolutionary usefulness of imitation in other situations.

Here we study the evolution of cooperation in spatial social dilemma games, but departing from the traditional assumption that strategies of players are the ones to potentially be imitated. Although it is certainly reasonable to assume that if one strategy is performing good imitating it is bound to yield positive results, we caution that this may not always be the case. Indeed, it is easy to come up with many such examples, the most obvious one being that imitating defection from a player that is surrounded by cooperators is a very bad idea if oneself is surrounded by defectors. Of course this scenario is more or less likely depending on the overlap between the neighborhoods of the two players, and may be more applicable to human societies than a grouping of simple microorganisms, yet it nevertheless is motivating enough for us to reconsider the concept of imitation. In particular, we refine it by allowing players not simply to imitate pure strategies, but rather to imitate emotional profiles of each other. In order to keep the model simple but still capturing the most relevant new features, we assign to every player two properties that define its emotional profile, namely the probability to cooperate with a more successful neighbor and the probability to cooperate with a less successful neighbor. With the first we determine envy or spite, while with the second property we determine goodwill or charity of each individual.
In this way the strategy becomes link-specific rather than player-specific, as is the case in the traditional model. Obviously, other interpretations of the two probabilities are possible as well. Interestingly, we find that, without any additional assumptions, by imitating the more successful emotional profiles instead of simply the more successful strategies, the evolution of cooperation is significantly promoted and substantially higher social welfare is attainable, even in games where the most favorable is the mixed strategy phase. Thus, a simple fine tuning of the concept of imitation, or rather of what is possible to imitate, reestablishes imitation as an important and globally beneficial behavior in evolutionary processes.

The remainder of this letter is organized as follows. First, we describe the considered social dilemmas and the protocol for the imitation of emotional profiles. Next we present the results, whereas lastly we summarize and discuss their implications.

\section{Social dilemmas and emotional profiles}

\begin{figure}
\scalebox{0.53}[0.53]{\includegraphics{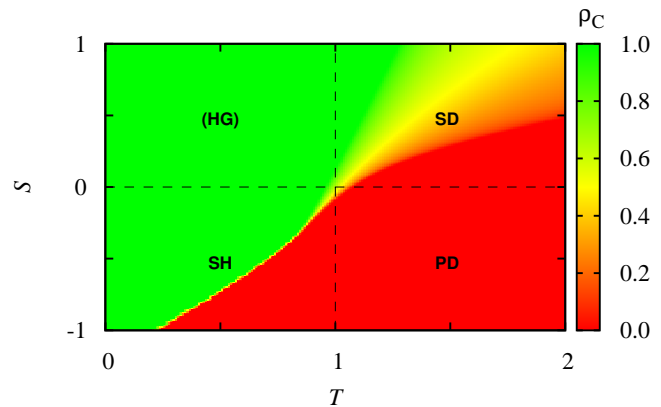}}
\caption{Schematic presentation of the two-dimensional $T-S$ parameter plane encompassing the stag-hunt (SH), the prisoner's dilemma (PD) and the snowdrift (SD) game. The upper left quadrant represents the so-called harmony game (HG), yet the latter does not constitute a social dilemma because there cooperation is always the winning strategy. The color mapping depicts the stationary fraction of cooperators $\rho_C$ as obtained if the evolutionary process is governed by strategy imitation. See also the main text for details.}
\label{traditional}
\end{figure}

Assuming that mutual cooperation yields the reward $R$, mutual defection leads to punishment $P$, and the mixed choice gives the cooperator the sucker's payoff $S$ and the defector the temptation $T$, we have the prisoner's dilemma game if $T>R>P>S$, the snowdrift game if $T>R>S>P$, and the stag-hunt game if $R>T>P>S$, thus covering all three major social dilemma types where players can choose between cooperation and defection. Following common practice, we set $R = 1$ and $P=0$, thus leaving the remaining two payoffs to occupy $-1 \leq S \leq 1$ and $0 \leq T \leq 2$, as depicted schematically in Fig.~\ref{traditional}.

In the traditional model, irrespective of the governing social dilemma, each player $x$ occupies a node on the $L \times L$ square lattice and is initially designated either as a cooperator $(s_x=C)$ or defector $(s_x=D)$ with equal probability, while evolution of the two strategies is performed in accordance with the Monte Carlo simulation procedure comprising the following elementary steps. First, a randomly selected player $x$ acquires its payoff $p_x$ by playing the game with all its four neighbors. Next, one randomly chosen neighbor of $x$, denoted by $y$, also acquires its payoff $p_y$ by playing the game with all its four neighbors. Traditionally player $y$ then imitates the strategy of player $x$ with the probability $q=1/\{1+\exp[(p_y-p_x)/K]\}$, where $K$ determines the level of uncertainty by strategy adoptions \cite{szabo_pre98}, which can be attributed to errors in judgment due to mistakes and external influences that affect the evaluation of the opponent. Without loss of generality we set $K=0.5$, implying that better performing players are readily imitated, but it is not impossible to adopt the strategy of a player performing worse. This value of $K$ is representative for a wide range of finite selection intensities. The weak-selection limit \cite{altrock_njp09, ohtsuki_jtb10,wu_b_pre10} ($K \to \infty$), however, is not studied in the present work. For this traditional setup the stationary fraction of cooperators $\rho_C$ in the $T-S$ parameter plane is as depicted in Fig.~\ref{traditional}. Well known results include the widespread dominance of defectors in the prisoner's dilemma quadrant, as well as the possibility of cooperator dominance and coexistence with the defectors in the snowdrift and the stag-hung quadrant, yet only for sufficiently favorable combinations of $T$ and $S$. These results will primarily be used for comparison purposes with the main findings that will be presented in the next section.

In order to depart from the traditional setup of spatial social dilemma games summarized above, we introduce an emotional profile to each player $x$, which is determined by the parameter pair $(\alpha_x, \beta_x) \in [0.1]$. Here $\alpha_x$ is the probability that player $x$ will cooperate with player $y$ if $p_x \geq p_y$, while $\beta_x$ is the probability that player $x$ will cooperate with player $y$ if $p_x < p_y$. Essentially thus, the two parameters determine how a given player will behave when facing more or a less successful opponent. Initially, to enable the start of the evaluation process, each player is assigned a random $(\alpha,\beta$) pair and a payoff from the reachable $[4S,4T]$ interval. Subsequently, every payoff value is updated by considering the proper neighborhoods of a player and the actual emotional parameters. Importantly, after the accumulation of new payoffs, player $y$ does not imitate the strategy of player $x$ with the previously established adoption probability $q$, but rather its emotional profile, \textit{i.e.} the $\alpha_x$ and/or $\beta_x$ value. Such a profile implicitly allows a player to behave differently (to cooperate and/or defect) towards different neighbors at the same time. Since the emotional profile consist of two parameters, however, the imitation is done separately for the two to avoid potential artificial propagation of freak (extremely successful) $(\alpha_x, \beta_x)$ pairs. Naturally, the same probability $q$ is applied for both imitations. Finally, after each imitation the payoff of player $y$ is updated using its new emotional profile, whereby each full Monte Carlo step involves all players having a chance to adopt the emotional profile from one of their neighbors once on average. Prior to presenting the result of this model, it is important to note that there will always be a fixation of $(\alpha_x, \beta_x)$ pairs, \textit{i.e.} irrespective of $T$ and $S$ only a single pair will eventually spread across the whole population. Naturally, the fixation time depends on the system size as well as game parametrization, which we have taken properly into account by sufficiently long simulations times prior to recording the final $\alpha$ and $\beta$ value. It is also worth pointing out that once the fixation occurs, the evolutionary process stops. The characteristic probability of encountering cooperative behavior on the spatial grid, which is equivalent to the stationary fraction of cooperators in the traditional version of the game, can thus be determined by means of averaging over the final states that emerge from different initial conditions.

\section{Results}

\begin{figure}
\scalebox{0.53}[0.53]{\includegraphics{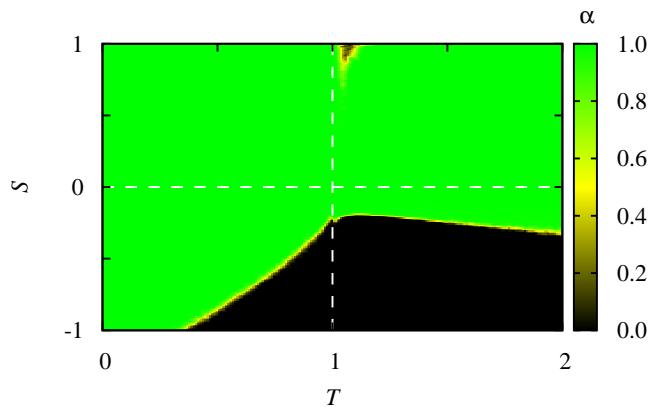}}
\caption{Color map depicting the final values of $\alpha$ on the $T-S$ parameter plane. Interpreting the probability to cooperate with less successful neighbors as goodwill or charity, it can be concluded that this emerges spontaneously for all three considered social dilemmas in at least some regions of the considered $T-S$ plane. Especially for the snowdrift quadrant (compare with Fig.~\ref{traditional}) the ``dominance'' of $\alpha=1$ is quite remarkable.}
\label{alpha}
\end{figure}

We start by presenting the color map encoding the final values of $\alpha$ on the $T-S$ parameter plane in Fig.~\ref{alpha}. Since $\alpha$ is the probability that players will cooperate with their less successful neighbors, \textit{i.e.} despite the fact that their payoff is lower, this can be interpreted either as goodwill or charity. From the presented results it follows that for the snowdrift quadrant this behavior is practically completely dominant, irrespective of the details of game parametrization. Thus, if the governing social dilemma is of the snowdrift type, then players will always ($\alpha=1$) cooperate with their neighbors provided their payoff is lower. For the stag-hunt game, on the other hand, the region of $\alpha=1$ corresponds roughly to the region of cooperator dominance in the traditional model (compare with results presented in Fig.~\ref{traditional}), although it extends somewhat further towards smaller $S$ and larger $T$ values. Results in the lower right quadrant, corresponding to the prisoner's dilemma game, are equally positive, indicating that as long as $S$ is not too low, cooperation with less successful neighbors will be the dominant behavior. This holds virtually independent of $T$, although surprisingly as $T$ increases the minimal $S$ still warranting $\alpha=1$ decreases. It can thus be concluded that raising the temptation to defect may even facilitate charitable actions in that they are upheld even by lower values of $S$.

\begin{figure}
\scalebox{0.53}[0.53]{\includegraphics{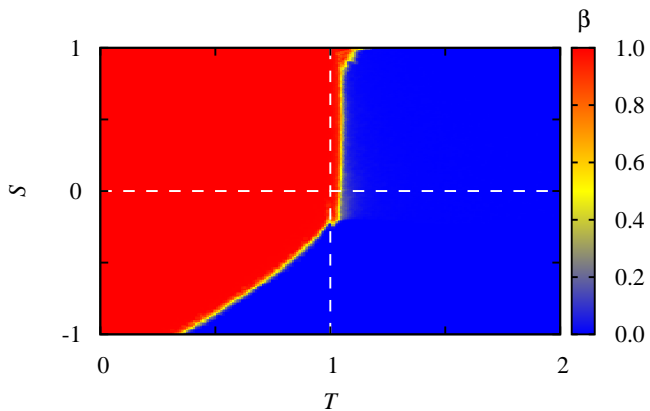}}
\caption{Color map depicting the final values of $\beta$ on the $T-S$ parameter plane. Interpreting the probability to cooperate with more successful neighbors as being representative for envy or spite, it can be concluded that this represent a serious impediment for the evolution of cooperation, especially for the snowdrift and the prisoner's dilemma game. There players will not cooperate with their neighbors if the latter are more successful than themselves. Conversely, in the stag-hunt game the payoff of neighbors, if compared to that of the player deciding either to cooperate or to defect, will play no role. Compare with results presented in Fig.~\ref{alpha}.}
\label{beta}
\end{figure}

Since the final values of $\alpha$ reveal only half of the behavior on the spatial gird, it is next of interest to examine the color map encoding the final values of $\beta$ on the $T-S$ parameter plane. Results presented in Fig.~\ref{beta} reveal at a glance that it is significantly more difficult to achieve cooperation with more successful neighbors than vice versa (compare with results presented in Fig.~\ref{alpha}). While the results for the stag-hung game for $\beta$ are practically identical to those for $\alpha$, the situation is much different for the snowdrift and the prisoner's dilemma game. In the snowdrift quadrant the total dominance of $\alpha=1$ is replaced by near complete dominance of $\beta=0$, indicating that players will not cooperate with their neighbors if the later are more successful. Envy thus appears to be an important agonist for the evolution of defection, rather than cooperation, in the snowdrift game. Only for values of $T$ slightly above $1$, and irrespective of $S$, will players choose to cooperate with their more successful neighbors, but otherwise not. For the prisoner's dilemma game the results are equally negative, further restricting cooperation with more successful neighbors not only to small values of $T$, but also only to moderately negative values of $S$. As we will reveal below, however, unwillingness to cooperate with the more successful neighbors has negative consequences mainly for the evolution of cooperation in the prisoner's dilemma game, while for the snowdrift game this fact actually favors the emergence of the globally optimal mixed phase warranting the highest level of social welfare.

\begin{figure}
\scalebox{0.53}[0.53]{\includegraphics{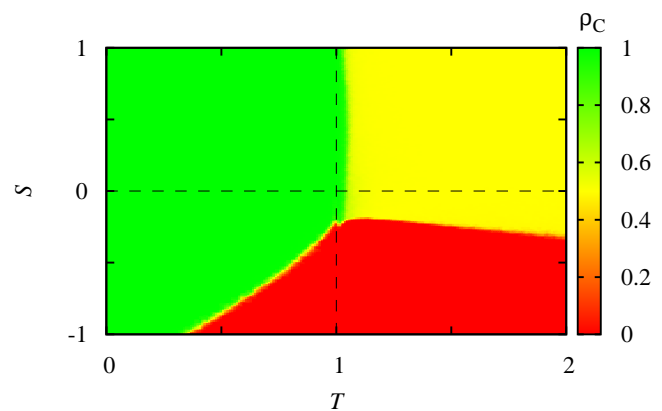}}
\caption{Color map depicting the final probability of cooperative behavior $\rho_C$ on the $T-S$ parameter plane. Since the probability to cooperate should be seen equal to the stationary fraction of cooperators in the traditional version of the game, a direct comparison of results presented in this figure and Fig.~\ref{traditional} clearly reveals that replacing the imitation of strategies with the imitation of emotional profiles, as determined by $\alpha$ and $\beta$, strongly promotes the evolution of cooperation in all three considered spatial social dilemma games.}
\label{cooperation}
\end{figure}

By considering the results presented in Figs.~\ref{alpha} and \ref{beta} combined, we arrive at the probability to encounter cooperative behavior on the spatial grid, as depicted in Fig.~\ref{cooperation}. Here $\rho_C$ denotes the average level of cooperative behavior on the spatial grid after the evolution of emotional profiles has stopped, \textit{i.e.} after the fixation of $\alpha$ and $\beta$. Since the regions of $\alpha=1$ and $\beta=1$ in the stag-hunt quadrant overlap completely, it is natural that in this region also the probability to encounter cooperation will be equal to $1$. Comparing this to the results presented in Fig.~\ref{traditional}, it can be concluded that replacing the imitation of strategies with the imitation of emotional profiles in the stag-hunt game promotes cooperation by extending the $\rho_C=1$ region towards larger values of $T$ and smaller values of $S$. For the snowdrift and the prisoner's dilemma game full dominance of cooperative behavior can be observed where $\alpha=1$ and $\beta=1$ regions overlap, while if $\alpha=1$ and $\beta=0$ the probability to encounter cooperative behavior equals $0.5$. Naturally, where both $\alpha$ and $\beta$ are equal to zero also $\rho_C=0$. Altogether, by comparing results presented in Figs.~\ref{traditional} and \ref{cooperation}, it can be concluded that imitating emotional profiles, and thus having the liberty to behave differently towards different players, instead of adopting pure strategies, strongly promotes the evolution of cooperation in all three considered spatial social dilemma games. Particularly players engaging in the snowdrift game profit immensely from the new imitation procedure, which is surprising since especially for spatial games having a mixed Nash equilibrium imitation has acquired quite a negative reputation \cite{hauert_n04}.

\begin{figure}
\centerline{\scalebox{0.30}[0.30]{\includegraphics{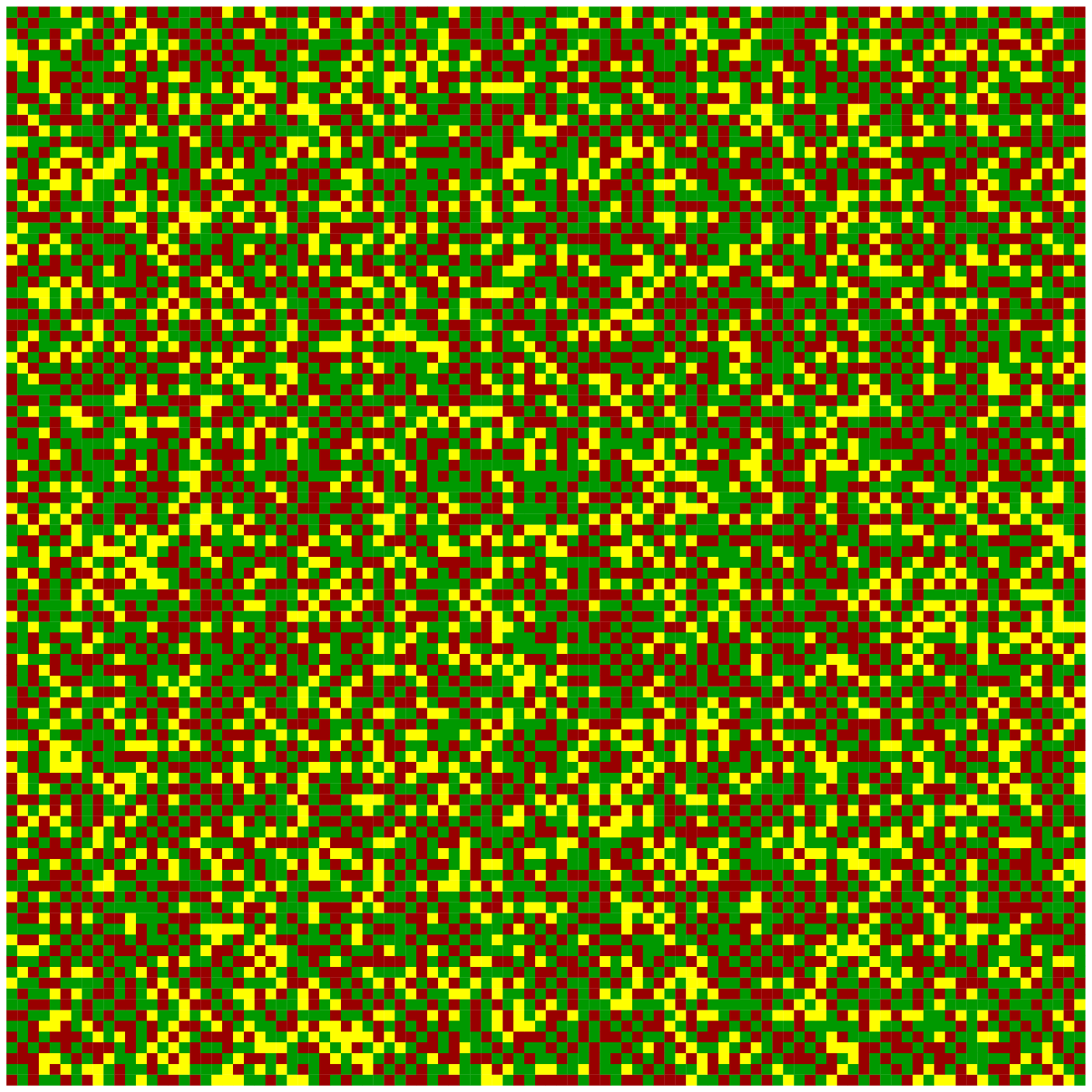}}
\scalebox{0.30}[0.30]{\includegraphics{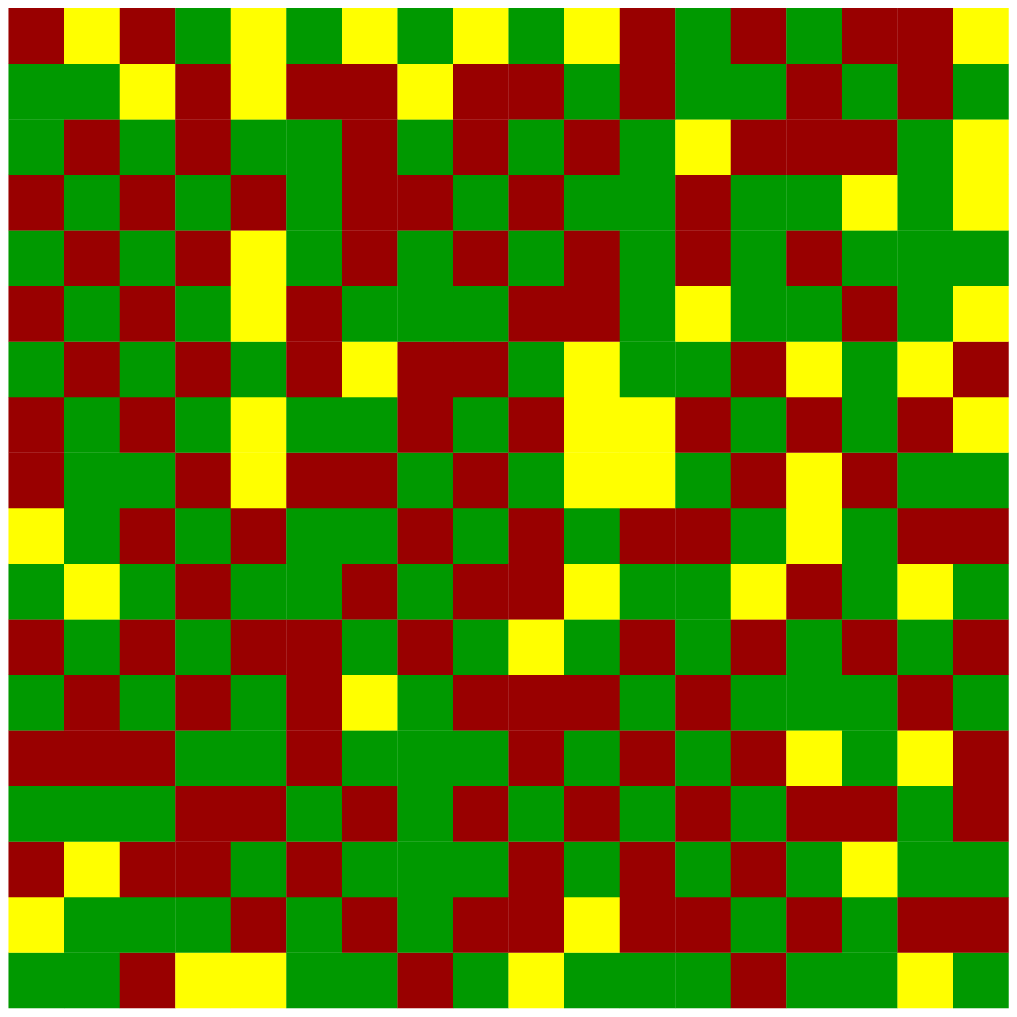}}}
\caption{Left: Characteristic snapshot of the final configuration of players engaging in the snowdrift game. Players are depicted green if they cooperate more frequently than defect with their neighbors, red if the opposite holds, and yellow if cooperation and defection are equally frequent. Right: Enlargement of a small portion of the spatial grid on the left (using the same color coding), revealing the characteristic role-separating checkerboard configuration of players, which warrants the highest mutual payoffs in the snowdrift game.}
\label{snap}
\end{figure}

\begin{figure}
\scalebox{0.53}[0.53]{\includegraphics{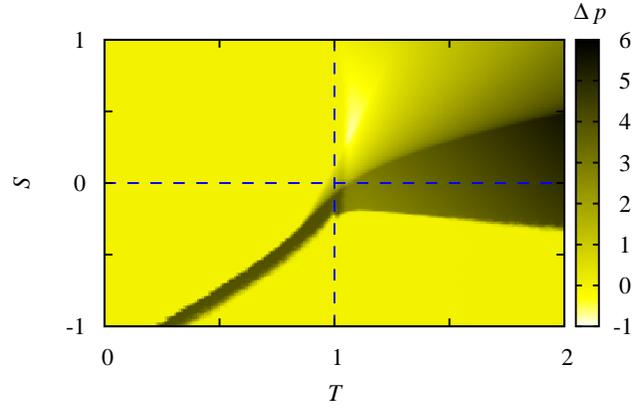}}
\caption{Color map depicting the difference in average payoffs $\Delta p$ between the traditional version of the considered spatial social dilemma games (results presented in Fig.~\ref{traditional}) and the new one adopting the imitation of emotional profiles instead of strategies. It can be observed that all three considered social dilemma games are able to benefit substantially from the updated form of imitation, increasing the social welfare by significant margins if compared to the traditional form of evolution that is governed by imitating player strategies. Quite unexpectedly, the advantages of updated imitation are most pronounced in the snowdrift quadrant (upper right), which due to the fact that the Nash equilibrium is a mixed phase, should be least susceptible to the benefits emerging as a result of imitation.}
\label{payoff}
\end{figure}

Since the success of imitation for spatial games where the Nash equilibrium is a mixed phase (coexistence of cooperators and defectors), as is the case for the snowdrift game, has often been questioned, it is thus of interest to examine results in this particular region of the $T-S$ parameter plane more precisely. Foremost, it should be emphasized that fine-tuning the imitation (what to imitate) procedure clearly restores the successfulness of imitation to arrive at a final state that is optimal for the society as a whole (see also results presented in Fig.~\ref{payoff} further below). The snapshot depicted in the left panel of Fig.~\ref{snap} presents a typical final configuration of players, color-coded in such a way that if the player behaves cooperatively more (less) frequently than defectively toward its neighbors it is depicted green (red), while if the two actions are equally frequent it is depicted yellow. The presented snapshot reveals a characteristic checkerboard distribution of expected strategies, which is made even clearer by the enlargement of a typical region of the spatial grid depicted in the right panel of Fig.~\ref{snap}. Noteworthy, as a result of the evolutionary process, and despite of diverse strategies, players exhibit identical willingness to either cooperate or to defect, \textit{i.e.} are characterized by the same emotional profile. This indicates that under the newly proposed imitation procedure players indeed share roles of cooperation and defection in order to arrive at the ``socially optimal'' configuration. Put differently, the spatial arrangement of players demonstrates that using the same attitude towards more or less successful players may result in a spatial mixture of cooperative and defective actions that warrants the highest mutual payoff.

Finally, as the last, and perhaps most persuading evidence for the successfulness of the newly proposed imitation procedure, it is thus instructive to examine the difference in payoffs between the traditional version of the games (results presented in Fig.~\ref{traditional}) and the one introduced here adopting imitation of emotional profiles instead of strategies. Results presented in Fig.~\ref{payoff} reveal most clearly the extent of cooperation promotion in the stag-hunt quadrant (the black stripe in the lower left quadrant corresponds accurately to the enlarged area of cooperator dominance), as well as the transition towards the socially optimal mixed $C-D$ phase in large regions of the snowdrift (upper right) quadrant. The prisoner's dilemma game, arguably constituting the most demanding conditions for the evolution of cooperation, also presents itself as very much susceptible to the positive impact of the new imitation procedure, if only the sucker's payoff is not too negative. Note also that the difference in the harmony quadrant and partly also in the stag-hunt quadrant is zero because both models yield a full $C$ phase. With these final results, we conclude that imitating emotions such as goodwill and envy instead of unconditionally copying pure strategies from the more successful players reestablishes imitation as perfectly suitable for resolving social dilemmas on structured populations, even for games where the Nash equilibrium is a mixed phase.

Before concluding, we note that the dynamics of the model proposed in this letter is significantly different from the one emerging when the evolutionary process is governed by stochastic reactive strategies \cite{nowak_aam90,szabo_pr07}. In the latter case, the choice of action in a given round is only affected by the opponent's behavior in the previous round, and consequently, a special form of reciprocity can emerge between neighbors because a cooperative act will likely trigger a similar reaction (to cooperate) from the targeted player. As we have argued, this is not necessarily true when emotions are subject to imitation. The role-separating mixed phase in the snowdrift quadrant has already been observed in spatial games, but it needed a significantly different -- the so-called myopic strategy update -- where a player can change the strategy independently from its neighborhood \cite{sysiaho_epjb05,szabo_pre10}. Results presented here reveal that such a state can evolve also by means of imitation. To highlight the robustness of our findings, we have also applied the so-called death-birth updating \cite{ohtsuki_jtb06}, but found very similar results. Furthermore, the application of weak mutation, allowing the emergence of independent $(\alpha,\beta)$ pairs, does not interfere with the evolution towards unique emotion profiles, as we have described above.

\section{Summary}

In sum, we have proposed and studied an alternative form of imitation, focusing specifically on its impact on the evolution of cooperation in the three most frequently considered spatial social dilemma games, namely the spatial snowdrift, stag-hunt and the prisoner's dilemma game. By replacing the imitation of strategies by the imitation of emotional profiles of players, as defined by the probability to cooperate with the more and less successful neighbors, we have found that players are much more likely to cooperate with less successful neighbors than they do with those who are more successful. Thus, while goodwill and charity appear to be important agonists facilitating the evolution of cooperation, envy and spite act detrimental, favoring the evolution of defection instead. Importantly, this duality in the evolution of the two emotional traits of players actually leads to rather unexpected benefits in the snowdrift game, where the Nash equilibrium is a mixed phase. Although imitation was previously thought to be unsuitable for achieving the socially optimal state in this type of spatial games, our results indicate that the limitations lie not in the act of imitation itself, but rather in what is available for imitation. By replacing the strategy with a slightly more elaborate concept of an emotional profile, we have found that imitation is fully capable of guiding the population towards the globally optimal state warranting the highest level of social welfare. The stag-hunt as well as the prisoner's dilemma game are also susceptible to the promotion of cooperation by means of the newly proposed imitation procedure. But while in the stag-hunt game benefits from both the cooperation with less as well as with the more successful neighbors are attainable, in the prisoner's dilemma game the positive impact on the evolution of cooperation is (almost) entirely due to players being willing to cooperate with their less successful neighbors. Envy, being prohibitive to act cooperatively with more successful neighbors, thus appears to be a major inhibitor of higher levels of cooperative behavior in the prisoner's dilemma game. Altogether, we find that more elaborate forms of imitation may reveal new mechanisms of promoting the evolution of cooperation in ways that appear to be more closely associated with complex societies, where the strategies alone may carry insufficient information to fully exploit the benefits of imitation.

\acknowledgments

Authors acknowledge support from the Hungarian National Research Fund (grant K-73449), the Bolyai Research Scholarship, the Natural Science Foundation of the Anhui Province of China (grant 11040606M119), and the Slovenian Research Agency (grants Z1-2032 and J1-4055).


\begin{thebibliography}{10}
\expandafter\ifx\csname url\endcsname\relax\def\url#1{\texttt{#1}}\fi

\bibitem{axelrod_84}
\Name{Axelrod R.} \Book{The Evolution of Cooperation} (Basic Books, New York)
  1984.

\bibitem{hofbauer_98}
\Name{Hofbauer J. \and Sigmund K.} \Book{Evolutionary Games and Population
  Dynamics} (Cambridge University Press, Cambridge, UK) 1998.

\bibitem{nowak_06}
\Name{Nowak M.~A.} \Book{Evolutionary Dynamics} (Harvard University Press,
  Cambridge, MA) 2006.

\bibitem{sigmund_10}
\Name{Sigmund K.} \Book{The Calculus of Selfishness} (Princeton University
  Press, Princeton, MA) 2010.

\bibitem{santos_pnas06}
\Name{Santos F.~C., Pacheco J.~M. \and Lenaerts T.} \REVIEW{Proc. Natl. Acad.
  Sci. USA}{103}{2006}{3490}.

\bibitem{hauert_jtb06b}
\Name{Hauert C.} \REVIEW{J. Theor. Biol.}{240}{2006}{627}.

\bibitem{fu_pla07}
\Name{Fu F., Chen X., Liu L. \and Wang L.} \REVIEW{Phys. Lett. A}{371}{2007}{58}.

\bibitem{gomez-gardenes_prl07}
\Name{G{\'o}mez-Garde{\~n}es J., Campillo M., Moreno Y. \and Flor{\' \i}a
  L.~M.} \REVIEW{Phys. Rev. Lett.}{98}{2007}{108103}.

\bibitem{rong_pre10}
\Name{Rong Z., Wu Z.-X. \and Wang W.-X.} \REVIEW{Phys. Rev. E}{82}{2010}{026101}.

\bibitem{tomassini_ijmpc07}
\Name{Tomassini M., Luthi L. \and Pestelacci E.} \REVIEW{Int. J. Mod. Phys. C}{18}{2007}{1173}.

\bibitem{szolnoki_epl09}
\Name{Szolnoki A. \and Perc M.} \REVIEW{EPL}{86}{2009}{30007}.

\bibitem{cremer_njp09}
\Name{Cremer J., Reichenbach T. \and Frey E.} \REVIEW{New J. Phys.}{11}{2009}{093029}.

\bibitem{poncela_epl09}
\Name{Poncela J., G{\'o}mez-Garde{\~n}es J., Flor{\' \i}a L.~M., Moreno Y. \and  S{\'a}nchez A.} \REVIEW{EPL}{88}{2009}{38003}.

\bibitem{dai_ql_njp10}
\Name{Dai Q., Li H., Cheng H., Li Y. \and Yang J.} \REVIEW{New J. Phys.}{12}{2010}{113015}.

\bibitem{helbing_pre10b}
\Name{Helbing D. \and Lozano S.} \REVIEW{Phys. Rev. E}{81}{2010}{057102}.

\bibitem{liu_rr_pa10}
\Name{Liu R.-R., Jia C.-X. \and Wang B.-H.} \REVIEW{Physica A}{389}{2010}{5719}.

\bibitem{lee_s_prl11}
\Name{Lee S., Holme P. \and Wu Z.-X.} \REVIEW{Phys. Rev. Lett.}{106}{2011}{028702}.

\bibitem{van-segbroeck_njp11}
\Name{Van~Segbroeck S., Santos F.~C., Lenaerts T. \and Pacheco J.~M.} \REVIEW{New J. Phys.}{3}{2011}{013007}.

\bibitem{nowak_n92b}
\Name{Nowak M.~A. \and May R.~M.} \REVIEW{Nature}{359}{1992}{826}.

\bibitem{hamilton_wd_jtb64a}
\Name{Hamilton W.~D.} \REVIEW{J. Theor. Biol.}{7}{1964}{1}.

\bibitem{axelrod_s81}
\Name{Axelrod R. \and Hamilton W.~D.} \REVIEW{Science}{211}{1981}{1390}.

\bibitem{wilson_ds_an77}
\Name{Wilson D.~S.} \REVIEW{Am. Nat.}{111}{1977}{157}.

\bibitem{traulsen_pnas06}
\Name{Traulsen A. \and Nowak M.~A.} \REVIEW{Proc. Natl. Acad. Sci. USA}{103}{2006}{10952}.

\bibitem{szolnoki_njp09}
\Name{Szolnoki A. \and Perc M.} \REVIEW{New J. Phys.}{11}{2009}{093033}.

\bibitem{nowak_s06}
\Name{Nowak M.~A.} \REVIEW{Science}{314}{2006}{1560}.

\bibitem{szabo_pr07}
\Name{Szab{\'o} G. \and F{\'a}th G.} \REVIEW{Phys. Rep.}{446}{2007}{97}.

\bibitem{schuster_jbp08}
\Name{Schuster S., Kreft J.-U., Schroeter A. \and Pfeiffer T.} \REVIEW{J. Biol.
  Phys.}{34}{2008}{1}.

\bibitem{roca_plr09}
\Name{Roca C.~P., Cuesta J.~A. \and S{\'a}nchez A.} \REVIEW{Phys. Life Rev.}{6}{2009}{208}.

\bibitem{perc_bs10}
\Name{Perc M. \and Szolnoki A.} \REVIEW{BioSystems}{99}{2010}{109}.

\bibitem{hauert_n04}
\Name{Hauert C. \and Doebeli M.} \REVIEW{Nature}{428}{2004}{643}.

\bibitem{grujic_pone10}
\Name{Gruji{\'c} J., Fosco C., Araujo L., Cuesta J.~A. \and S{\'a}nchez A.}
  \REVIEW{PLoS ONE}{5}{2010}{e13749}.

\bibitem{cook_prsb11}
\Name{Cook R., Bird G., L{\"u}nser G., Huck S. \and Heyes C.} \REVIEW{Proc. R.
  Soc. B}{}{2011}{}.

\bibitem{szabo_pre98}
\Name{Szab{\'o} G. \and T{\H{o}}ke C.} \REVIEW{Phys. Rev. E}{58}{1998}{69}.

\bibitem{altrock_njp09}
\Name{Altrock P.~M. \and Traulsen A.} \REVIEW{New J. Phys.}{11}{2009}{013012}.

\bibitem{ohtsuki_jtb10}
\Name{Ohtsuki H.} \REVIEW{J. Theor. Biol.}{264}{2010}{136}.

\bibitem{wu_b_pre10}
\Name{Wu B., Altrock P.~M., Wang L. \and Traulsen A.} \REVIEW{Phys. Rev. E}{82}{2010}{046106}.

\bibitem{nowak_aam90}
\Name{Nowak M. \and Sigmund K.} \REVIEW{Acta Appl. Math.}{20}{1990}{247}.

\bibitem{sysiaho_epjb05}
\Name{Sysi-Aho M., Saram{\"a}ki J., Kert{\'e}sz J. \and Kaski K.} \REVIEW{Eur.
  Phys. J. B}{44}{2005}{129}.

\bibitem{szabo_pre10}
\Name{Szab{\'o} G., Szolnoki A., Varga M. \and Hanusovszky L.} \REVIEW{Phys.
  Rev. E}{80}{2010}{026110}.

\bibitem{ohtsuki_jtb06}
\Name{Ohtsuki H. \and Nowak M.~A.} \REVIEW{J. Theor. Biol.}{243}{2006}{86}.

\end{thebibliography}
\end{document}